\documentclass[pre,groupedaddress,twocolumn,a4paper,showpacs]{revtex4-1}
\usepackage{graphics}
\usepackage{amssymb}
\usepackage{amsmath}
\usepackage{wasysym}
\usepackage{latexsym}
\usepackage{graphicx}
\usepackage{epsfig}
\usepackage{subfigure}
\usepackage{float}
\usepackage{color}
\usepackage{lineno}
\usepackage{verbatim}

\begin{document}


\title{The dynamics of casual groups can keep free-riders at bay}

\author{Jos\'e F.  Fontanari}
\affiliation{Instituto de F\'{\i}sica de S\~ao Carlos, Universidade de S\~ao Paulo,  13560-970 S\~ao Carlos, S\~ao Paulo, Brazil}
  
 \author{Mauro Santos}
\affiliation{Departament de Gen\`etica i de Microbiologia, Grup de Gen\`omica, Bioinform\`atica i Biologia Evolutiva (GBBE), Universitat Aut\`onoma de Barcelona, Spain \\
cE3c - Centre for Ecology, Evolution and Environmental Changes \& CHANGE - Global Change and Sustainability Institute, Lisboa, Portugal}

\begin{abstract}
Understanding the conditions for maintaining cooperation in groups of unrelated individuals despite the presence of non-cooperative members is a major research topic in contemporary biological, sociological, and economic theory. The $N$-person snowdrift game models the type of social dilemma where cooperative actions are costly, but there is  a reward for performing them. We study this game in a scenario where players move between  play groups following the casual group dynamics, where groups grow by recruiting isolates and shrink by losing individuals who then become isolates. This describes the size distribution of spontaneous human groups and also the formation of sleeping groups in monkeys. We consider three scenarios according to the probability of  isolates joining a group. We find that for appropriate choices of the cost-benefit ratio of cooperation and the aggregation-disaggregation ratio in the formation of casual groups, free-riders can be completely eliminated from the population. If individuals are more attracted to large groups, we find that cooperators persist in the population even when the mean group size diverges. We also point out the remarkable similarity between the replicator equation approach to public goods games and the trait group formulation of structured demes.
\end{abstract}

%

\maketitle

\section{Introduction}\label{sec:intro}

The problem of cooperation in social dilemmas is that individuals face conflicting interests: either they incur costs to cooperate and further the collective interest, or free ride on the fruitful efforts of others to pursue their own interest. The traditional game theoretical framework for analyzing this problem is the prisoner's dilemma for pairwise interactions or, more generally, the public goods $N$-person prisoner's dilemma \cite{Peterson_2015}. It describes situations in which the costly cooperation performed by a focal individual only increases others' benefits. In this case, it is well known that in randomly formed groups of  $N \geq 2$ individuals that
last less than one generation \cite{Fletcher_2004}, cooperation is doomed to extinction \cite{Hamilton_1975}, and that the evolution and maintenance of cooperators requires some positive assortment between them, which is best illustrated by the green-beard effect \cite{Hamilton_1964}, or repeated interactions   \cite{Axelrod_1984}  that allow the emergence of reputation and reciprocity \cite{Wang_2023}. However, it has been argued that the ubiquity of cooperation in nature, from populations
of microbes to organisms gifted with complex cognition, is likely not due to these mechanisms, but rather to nonlinearities in the benefits or costs accruing to players in a public goods game \cite{Archetti_2012}. The $N$-person prisoner's dilemma, where the reward to cooperators increases linearly with their number in the play group, requires all of these mechanisms (and more \cite{Doebeli_2005}) to ensure the maintenance of cooperators in the population (see, e.g., \cite{Hannelore_2010,Sigmund_2010}). 

There are many situations where the focal individual also benefits from its own cooperative strategy \cite{Archetti_2012}. It is worth recalling that John Maynard Smith, one of the founders of evolutionary game theory, used the hawk-dove game to introduce the concept of evolutionarily stable strategy \cite{Maynard_1982}. This game is equivalent to the chick\-en or $2$-person snowdrift game, in which two drivers are stuck on either side of a snowdrift blocking the road. Each driver has the option of either  to get out of the car and shovel (cooperate), or to stay in the car and let the other do the job (defect). 
A cooperator's payoff is $R=b -c/2$ when she plays against another cooperator and $S=b-c$ when she plays against a defector, while a defector's payoff is $T=b$ when she plays against a cooperator and $P=0$ when she plays against another defector. Here $b$ is the benefit of clearing the road and  $c$ is the cost of this task, where $b > c$, so $T>R>S>P$. If $b<c$,  the best strategy is to defect no matter what the other does, and we are back to the prisoner's dilemma with $T>R>P>S$.
The central idea of the snowdrift  game is that each player's best choice depends on what the other player chooses: it is better to shovel when the other player defects, and to defect when the other player shovels. This maintains cooperation in a mixed-stable state when the payoff of being stuck (mutual defection) is less than the payoff of clearing the road and going home \cite{Maynard_1982,Maynard_1973}. 
More precisely, this is only true if the game is symmetric and the players do not choose their strategies sequentially. If they choose their strategies sequentially, then the strategies where the first player defects and the second cooperates, or where the first cooperates and the second defects, are the evolutionarily stable solution of the game \cite{Kun_2006}.

Although $2$-person games, in which a player's fitness is determined by the cumulative reward from a given number of pairwise interactions with the other players, have been used to study cooperation, the social dilemmas that motivated these studies actually involved the interaction of  large numbers of individuals with no discernible pairwise interactions \cite{Hardin_1968,Ostrom_1990}, which most likely 
cannot be  reduced  to pairwise interactions \cite{Perc_2013}.  In contrast to the $N$-person prisoner's dilemma, the $N$-person snowdrift game, where the cost of cooperation decreases as the reciprocal of the number of cooperators (i.e., they share the cost of completing the task), exhibits coexistence between cooperators and defectors for finite $N$ without any ad hoc mechanism  favoring cooperation \cite{Zheng_2007}. Interestingly, in social psychology, the disappearance of cooperators as $N$ increases is explained by the diffusion of responsibility effect: people often fail to cooperate because they believe that others will or should take responsibility for helping a person in need \cite{Kassin_2013}. We will not address this social psychological issue here.

A significant advance in the field of public goods games is the realization that the replicator equation formalism used to study the biological evolutionary process in continuous time  \cite{Hofbauer_1998} also describes the cultural evolutionary process where individuals imitate the behavior of their more successful peers \cite{Traulsen_2005} (see also \cite{Sandholm_2010}). Here we ask the question of how to promote the increase of the number of cooperators and thus the average fitness of the population in the $N$-person snowdrift game in the context of free-forming or casual groups \cite{Coleman_1961,White_1962}, in which individuals are in face-to-face interactions and isolated players are free to join groups at a rate of $\lambda$, or leave groups to become isolates at a rate of $\mu$. Besides describing the size distribution of spontaneous human groups \cite{Coleman_1961}, casual group dynamics explains the formation of sleeping groups in monkeys, where some sort of public goods games are likely to be played \cite{Cohen_1971}. The attractiveness of a group to isolates is proportional to $N^\alpha$, where $N$ is the number of players in the play group \cite{Fontanari_2023}.  Large negative values of $\alpha$ describe situations where a predominance of pairs and isolates is expected, while large positive values of $\alpha$ favor the formation of a single large group coexisting with isolates.

The snowdrift game is played continuously as the group sizes and compositions change following the casual group dynamics, but it is assumed that the group dynamics is much faster than the imitation dynamics that determines the frequencies of the players' strategies in the infinite population. Thus, the sizes of the play groups are effectively distributed by the equilibrium group-size distribution of the casual group dynamics.  In particular, these distributions are  the zero-truncated Poisson ($\alpha=0$), the logarithmic series ($\alpha=1$), and the isolate-pair Bernoulli  ($\alpha \to -\infty$). These equilibrium distributions are parameterized only by the ratio of the aggregation and disaggregation rates. Using the framework of the replicator equation with the fitness of the players averaged over the group sizes, we find that the diversity of group sizes produced by the group dynamics promotes cooperation, and for appropriate choices of the cost-benefit ratio of  cooperation, the defectors can be completely eliminated from the population. Surprisingly, for the  logarithmic series distribution, but not for the zero-truncated Poisson distribution,  we find that the cooperators persist in the population  even when the mean group size diverges, a scenario that  occurs when $\mu \to 0$. In contrast, for play groups of fixed size $N$, the fraction of cooperators  in the infinite population vanishes as $1/N$  when  $N$ diverges \cite{Zheng_2007}. In many respects, our paper extends and complements previous  results on  snowdrift games with random groupings \cite{Xu_2022},  where  the group size $N$ is chosen at random  among the integers $\{1,2, \ldots, N_m \}$.  Although this uniform distribution cannot be obtained from the casual group dynamics,  the presence of  isolates (i.e., groups of size $N=1$) guarantees the existence of an all-cooperators regime.  In this uniform group size scenario, the fraction of cooperators vanishes as the upper bound $N_m$ diverges.

An interesting point that seems to have been largely overlooked in the evolutionary game literature is that the replicator equation approach to infinite population public goods games (see, e.g., \cite{Hannelore_2010,Zheng_2007}) is virtually identical to the trait group framework proposed by Wilson in the  1970s \cite{Wilson_1975,Wilson_1980,Okasha_2009}. In fact, in Wilson's trait group formulation, the fitness of individuals is determined locally within their trait groups, but the competition for reproduction involves the entire population. In the replicator equation formulation of the $N$-person public goods games, individuals' rewards are obtained by playing in groups of $N$ players, but the individual chosen as a model for imitation is chosen from the population at large. In particular, we discuss  how  the increase in variability of  the cooperator-defector ratio in the trait-groups, promoted by the diversity of group sizes, strengthens  cooperation as predicted by Wilson's trait group formulation.

The rest of the paper is organized as follows.   In Section \ref{sec:game}, we present the $N$-person snowdrift game and describe its formulation in the framework of the replicator equation, which is valid for an infinite population of players.  We discuss the equilibrium solutions of the replicator equation in the case where the play groups are formed by drawing a fixed number $N$ of players from the infinite population \cite{Zheng_2007}.
 In Section \ref{sec:casual} we describe the system of casual or freely forming groups,  in which individuals are free to maintain or break  contact with each other  \cite{Coleman_1961,White_1962,Cohen_1971,Fontanari_2023}.  Groups  grow by recruiting isolates and shrink by losing individuals who then become isolates.
 In Section \ref{sec:varN} we discuss how to obtain the  equilibrium cooperator  frequencies when  the group dynamics is much faster than the imitation dynamics,  resulting in an annealed average over the equilibrium group-size distributions. In Section \ref{sec:res} we discuss the equilibrium solutions of the replicator equation for the zero-truncated Poisson, the logarithmic series and the isolate-pair Bernoulli group-size distributions. 
Finally, in Section \ref{sec:conc} we discuss our results in the light of Wilson's trait group framework and compare our approach and results with previous work on  the  snowdrift games with dynamic groupings.

\section{The $N$-person snowdrift game}\label{sec:game}

In the context of the evolution of cooperation \cite{Axelrod_1984},  the $2$-person snowdrift game differs from the $2$-person prisoner's dilemma in that there is stable coexistence between cooperators and defectors if the payoff of mutual defection (P) is less  than the payoff of cooperating against a defector (S) \cite{Maynard_1982}. Now, the question is how to promote the increase of the number of cooperators, and thus the average fitness of the population \cite{Archetti_2012}, in the public goods   $N$-person snowdrift game \cite{Zheng_2007},  where the play group  consists of $N$ individuals with $i$ cooperators and $N-i$ defectors, so that the payoff $P_c(i)$ for a cooperator is
\begin{equation}\label{Pci0}
P_c(i) = b - \frac{c}{i},
\end{equation}
for $i=1, \dots,N$, while the payoff for a defector is $P_d(i) = b $ for $i=1, \dots,N-1$ and $P_d(0) = 0 $.  The non-linear dependence of cooperators' payoffs on the number of cooperators $i$ makes it impossible to describe the game as a combination of pairwise interactions between players \cite{Archetti_2012}.  Some interesting variants of the $N$-person snowdrift game assume that the task can only be completed if the number of cooperators in a group is above a threshold \cite{Pacheco_2009}, or that faster completion of the task leads to better payoffs for all players in the group \cite{Ji_2010}.

For play groups of fixed size formed by  randomly drawing   $N$  players from an infinite population where the proportion of cooperators is $x \in [0,1]$, the  average payoff of a cooperator is \cite{Zheng_2007}
\begin{eqnarray}\label{fc1}
f_c (N,x)  & = &  \sum_{j=0}^{N-1} \binom{N-1}{j} x^j (1-x)^{N-1-j} P_c(j+1) \nonumber \\
& = &  b - \frac{c}{x}  \frac{1}{N} \left [ 1 -( 1-x)^N  \right ],
\end{eqnarray}
and the  average payoff of a defector is 
\begin{eqnarray}\label{fd1}
f_d (N,x)  & = &  \sum_{j=0}^{N-1} \binom{N-1}{j} x^j (1-x)^{N-1-j} P_d(j) \nonumber \\
& = &  b \left [ 1 -(1-x)^{N-1} \right ] .
\end{eqnarray}

We assume that social learning, which is a wide\-spread strategy in nature \cite{Rendell_2010}, determines the strategies of players, who tend to imitate successful individuals and thus change their strategies accordingly. Explicitly, a randomly chosen player compares her average payoff with that of another randomly chosen player in the population, and adopts the strategy of the other player with a probability proportional to the payoff difference if it is positive. Otherwise, she keeps her own strategy. 

For an infinite well-mixed population, the frequency $x$ of cooperators is given by the replicator equation  \cite{Traulsen_2005}
\begin{eqnarray}
\dot{x}  & =  & x \left [ f_c (N,x) - \bar{f} \right ] \nonumber \\
& =  & x(1-x) \left [ f_c (N,x) - f_d (N,x) \right ] , \label{rep1}
\end{eqnarray}
where $\bar{f} = x f_c + (1-x) f_d $ is the average payoff of the entire population. We refer the reader to Ref. \cite{Hofbauer_1998} for a thorough introduction to the replicator equation.  The average payoff can be  written explicitly as a  function of  $N$ and $x$, viz.,  
\begin{equation}\label{barf}
\bar{f} (N,x) = \left ( b - \frac{c}{N} \right ) \left [ 1 - (1-x)^N \right ] ,
\end{equation}
which shows that $\bar{f}$  increases monotonically from $0$ to $b-c/N$ as $x$ increases from $0$ to $1$.  In this sense, promoting an increase in the equilibrium  frequency  of cooperators  amounts to promoting an  increase in  the average payoff of the population.  
 
Equation (\ref{rep1})  has three fixed-points, viz., $x=0$, $x=1$ and 
$x=x_N^*$, where $x_N^*$ is the solution of the $N$-th-order polynomial  equation  \cite{Zheng_2007}
\begin{equation}\label{fc=fd}
f_c(N,x_N^*) = f_d(N,x_N^*) .
\end{equation}
Here the subscript $N$ reminds us that the coexistence fixed point is for play groups of fixed size $N$.  For instance, $x_1^* =1$ and 
\begin{equation}\label{x2}
x_2^* = \frac{1-r}{1-r/2},
\end{equation}
where  $r=c/b$. We recall that for $2$-person games,  $ r<1$ corresponds to the  snowdrift  game, and $r>1$  to the prisoner's dilemma.
The standard local stability analysis \cite{Strogatz_2014} shows that the all-defectors fixed point $x=0$ is unstable when $r<1$, regardless of the value of $N$.  Similarly, the all-cooperators fixed point $x=1$ is always unstable  if the task is costly to complete since $ -f_c(N,1) + f_d(N,1) =c/N > 0$. Thus, for $b>c>0$, the coexistence fixed point $x_N^*$ is the only stable fixed point of the replicator equation (\ref{rep1}).  For $N \to \infty$, equation (\ref{fc=fd}) implies that $x_N^* \to 0$. More explicitly, 
for $N \to \infty$ and $x_N^* \to 0$ such that  $y = N x_N^*$ is finite and nonzero, Eq. (\ref{fc=fd})  is rewritten as 
\begin{equation}\label{y1}
r =  \frac{y}{e^{y}-1} .
\end{equation}
Noticing that the function on the right hand side of Eq. (\ref{y1}) decreases monotonically from $1$ to $0$ as $y$ increases from $0$ to $\infty$, we conclude that Eq. (\ref{y1}) has a unique solution for all $r<1$. In particular, for $r \approx 1$ we find 
\begin{equation}\label{y2}
y = 2 (1-r).
\end{equation}
This analysis shows that $x^*_N = y/N$ vanishes with the inverse of the group size $N$ for large $N$. We note that Eq. (\ref{y1})  gives the correct solution for the coexistence fixed point in the limit of large $N$ for an arbitrary value of $r<1$, while the equation offered in Ref. \cite{Zheng_2007} is only valid  for $r \approx 1$ and is equivalent to  Eq. (\ref{y2}).

To conclude this overview of the $N$-person snowdrift game,  we write the solution of Eq. (\ref{fc=fd}) for $x_N^* \approx 0$, which takes place for $r \approx 1$, viz., 
\begin{equation}\label{xN0}
x_N^* = 2 \frac{1-r}{N-1}
\end{equation}
and is valid for all $N> 1$.

\section{Casual group dynamics}\label{sec:casual}

We consider a fixed  large number of individuals $M$ in a closed system, which organize themselves into a variable number $K$ of groups 
of size $N=1, \ldots, M$.  These are the finite size play groups  introduced in Section \ref{sec:game}. If we denote by $n_N$ the number of groups of size $N$, then we have the constraints $\sum_{N=1}^M N n_N = M$ and $\sum_{N=1}^M n_N = K$. Both $n_N$ and $K$ are variables  determined by the casual group dynamics described next. We will  eventually let $M \to \infty$  to get  the infinite population needed for the validity of the replicator equation framework. Note that $M$ has no counterpart in the $N$-person public goods games discussed before, which assumes an infinite population from the start. Our focus is on the mean fraction of groups of size $N$, i.e., the group-size distribution
\begin{equation}
p_N = \frac{n_N}{K}
\end{equation}
with $N=1, \ldots, M$. For certain values of the attractiveness parameter $\alpha$, $p_N$ can be expressed in closed analytic  form in the limit $M \to \infty$ \cite{Fontanari_2023}.

 The process of joining and leaving the groups is as follows \cite{Coleman_1961,White_1962}. 
Each individual in a group of size $N> 1$ has the probability $\mu \delta t$ of leaving the group during the time interval $\delta t$. When an individual leaves a group, it becomes an isolate, i.e., a group of size $N=1$. An isolate has a probability $\lambda \delta t$ of joining a group (including other isolates) in the time interval $\delta t$. The attractiveness to isolates of a group of size $N$ is proportional to $N^\alpha$. The case $\alpha = 0$ describes the situation where isolates join every group in the system at the same rate regardless of its size \cite{Coleman_1961}.  For $\alpha > 0$ we have a contagious scenario that favors the formation of large groups \cite{White_1962}, while for $\alpha < 0$ we have an aversive scenario that discourages the formation of  large groups \cite{Fontanari_2023}. What makes this model amenable to an analytical solution is that the possibility of a group of size $L \geq 2$ joining or leaving a group of size $N \geq 2$ is ignored (see \cite{Okubo_1986} for a more general formulation of the  clustering-splitting process).

Note that the mathematical modeling of casual groups  initiated in the 1960s is based on the admittedly extreme assumption  that all individuals are equivalent:  ``wipe blank any prior knowledge of the individuals present''  \cite{White_1962}. Thus, the only perceptible difference between groups is their size $N$, and consequently, the attractiveness of a group to isolates must be a function of $N$ only.  As already noted, the particular choice of $N^\alpha$ has the advantage of recovering the two well-studied cases $\alpha=0$ and $\alpha=1$, as well as allowing the study of the extreme aversion ($\alpha \to -\infty$) and attraction ($\alpha \to \infty$)  to large groups \cite{Fontanari_2023}.   Casual groups are the short-lived groups that form within stable social groups and, unlike stable groups \cite{Dunbar_1992}, do not appear to have an optimal size \cite{Coleman_1961}.

In the limit $M \to \infty$, the equilibrium group-size distribution is the zero-truncated Poisson distribution for $\alpha =0$ and the logarithmic series distribution for $\alpha=1$. For $\alpha \to -\infty$ we have a ballroom scenario where only isolates and pairs are present
(see \cite{Fontanari_2021}  for an application of the pairs and isolates scenario to model loneliness).
We will consider only these three cases here.  There are no theoretical (analytical or numerical) results  for  $\alpha>1$, since  the mean-field approximation, which allows  the exact calculation of the group size distribution in the infinite population limit, breaks down. The only alternative  in this case  is to use  Gillespie’s stochastic algorithm to simulate the casual group dynamics directly \cite{Fontanari_2023}. 
Since the calculations of the equilibrium group-size distributions have recently been revised \cite{Fontanari_2023}, only the final results are presented below. These distributions depend only on the ratio between the aggregation and disaggregation rates, $\kappa = \lambda/\mu$.

\subsection{Zero-truncated Poisson distribution}

If isolates join groups regardless of their size (i.e., $\alpha =0$), the equilibrium group-size distribution in the infinite population limit is \cite{White_1962,Fontanari_2023}
\begin{equation}\label{pi_a0}
p_N = \frac{1}{\kappa} \frac{ \left [ \ln ( 1 + \kappa) \right ]^N}{N!} ,
\end{equation}
which we identify as the zero-truncated Poisson distribution. Interestingly, this distribution fits a wide variety of small group data, such as pedestrians on a sidewalk, playgroups on a playground, and shopping groups \cite{Coleman_1961}  (see \cite{Burgess_1984,Cattuto_2010,Starnini_2016} for  more recent empirical studies of   human casual groups).
The mean group size  is
\begin{equation}\label{i_al0}
m_{P}  = \sum_{N=1}^\infty N p_N  =  (1 +1/\kappa) \ln ( 1 + \kappa ),
\end{equation}
which  is approximately $1+ \kappa/2$ for small $\kappa$ and diverges very slowly with increasing  $\kappa$.

\subsection{Logarithmic series distribution}

If the  attractiveness  of groups to isolates  grows linearly with their size (i.e., $\alpha=1$), the equilibrium group-size distribution in the infinite population limit is \cite{White_1962,Fontanari_2023}
\begin{equation}\label{pi_a1}
p_N =  \frac{1}{\ln ( 1 +\kappa) } \frac{1}{N} \left ( \frac{\kappa}{1 + \kappa} \right )^N,
\end{equation}
which we identify as the logarithmic distribution used to   model relative species abundance \cite{Fisher_1943}. Thus, the mean group size
is 
\begin{equation}\label{i_al1}
m_{L}   =  \frac{\kappa}{ \ln ( 1 + \kappa )}
\end{equation}
which  is approximately $1+ \kappa/2$ for small $\kappa$ as for the zero-truncated Poisson distribution.

\subsection{Isolate-pair Bernoulli distribution}

If isolates are only attracted to other isolates (i.e., $\alpha \to - \infty$), then the resulting system is composed of isolates and pairs in the equilibrium regime. The fractions of isolates and pairs are  \cite{Fontanari_2023}
\begin{eqnarray}
p_1  & =  & \frac{1}{1 + \kappa/2}  \label{p1i} \\
p_2 & = &  \frac{\kappa/2}{1 + \kappa/2} \label{p2i}
\end{eqnarray}
with $p_1+p_2 =1$. The mean group size is
\begin{equation}\label{i_ali}
m_{B} = \frac{1+\kappa}{1+\kappa/2} ,
\end{equation}
which is approximately $1+\kappa/2$ in the small $\kappa$ limit as for the previous distributions.  Thus, in this limit, individuals are either isolated or form pairs, regardless of the form of the group attraction, i.e., regardless of the value of $\alpha$. 

 \section{Snowdrift in casual groups}\label{sec:varN}
 
Here we consider a scenario where the game takes place among the members of the casual groups. We assume that the time scale of the aggregation-disaggregation process is much faster than the imitation time scale, so that the players accumulate the payoffs received in infinitely many games taking place in a variety of group sizes (and compositions) determined by the casual group dynamics. 
The assumption that the casual group dynamics is much faster than the imitation dynamics is crucial because it allows us to use the equilibrium group size distribution $p_N$ of casual group dynamics in evaluating the average fitness.  In other words, players move between isolates and groups of size $N>1$ and back so quickly that equilibrium is reached almost instantaneously.
Thus the  average payoffs of a cooperator and of a defector in an infinite population with a proportion $x$ of cooperators are
\begin{equation}\label{Fc}
F_c(x) = \sum_{N=1}^\infty p_N f_c (N,x)
\end{equation}
and
\begin{equation}\label{Fd}
F_d(x) = \sum_{N=1}^\infty p_N f_d (N,x),
\end{equation}
where $f_c(x,N)$ and $f_d(x,N)$ are given by Eqs. (\ref{fc1}) and (\ref{fd1}), respectively, and $p_N$ is the equilibrium group-size distribution of the casual group dynamics.  The weighted sum over the different group sizes $N$ in these equations implies that players move from one group to another while playing the snowdrift game with fixed $x$. In the statistical physics of disordered systems, this is known as the annealed average and involves averaging over fast random variables (here the group sizes and compositions) while keeping the cooperator frequency $x$ fixed \cite{Mezard_1987}.

Note that in the replicator equation approach to public goods games, the imitation timescale is much slower than any timescale of the system.  This assumption is rarely made explicit, but it is necessary, for example, to write Eqs. (\ref{fc1}) and (\ref{fd1}) for fixed $N$, which  imply that players play infinitely many games for a fixed cooperator frequency $x$,  before engaging in the imitation process that would change $x$.
It is the slowness of the imitation dynamics that justifies replacing a player's payoff by her average payoff, as done in Eqs. (2) and (3). Thus, the additional averaging over $N$ described in Eqs. (\ref{Fc}) and (\ref{Fd}) is not a big leap from what is done in the case of fixed $N$.
Although this assumption may not be biologically or sociologically plausible, it allows the full use of the replicator equation approach and, somewhat surprisingly,  does not seem to affect the equilibrium regime of the dynamics, as indicated by the good agreement between the simulations where a player's payoff is not replaced by her average payoff and the analytical results of the replicator equation \cite{Zheng_2007}. Finding  a more robust mathematical argument to justify this agreement  is an interesting open research problem in evolutionary game theory.

Here we study the equilibrium solutions of the annealed replicator equation
\begin{equation}\label{xann}
\dot{x}_{ann} =x_{ann}(1-x_{ann}) \left [ F_c (x_{ann}) - F_d (x_{ann}) \right ],
\end{equation}
and show that the group dynamics stabilizes the all-cooperators fixed point $x_{ann} = 1$. The same result has been obtained using a uniform distribution of group sizes and considering a scenario in which both the group and imitation dynamics have the same time scale \cite{Xu_2022}. This choice of timescales precludes the use of the replicator equation formalism, which, as we will see next, greatly simplifies the mathematical analysis. In addition to the equilibrium solutions $x_{ann}^*$ of Eq. (\ref{xann}), we also consider the equilibrium frequency of cooperators for  a group size fixed at the mean value $m = \sum_{N=1}^\infty Np_N $. Using Jensen's inequality, we can prove that $x_{ann}^* \geq x_m^*$  under quite general conditions, i.e., $f_c(N,x) - f_d(N,x)$ is a strictly convex function of $N$ for all $x$ \cite{Pena_2011}.

Also in the case of variable group size, the population average payoff 
\begin{equation}
\bar{F}(x_{ann} ) =  \sum_{N=1}^\infty p_N  \bar{f} (N,x_{ann}) ,
\end{equation}
where $\bar{f} (N,x_{ann})$ is given by Eq. (\ref{barf}),  is a monotonically increasing function of $x_{ann}$. Thus, increasing the equilibrium frequency of cooperators leads to an increase in the average payoff of the population, which is a measure of public goods.

A different  annealed average has been used to study the effect of group size diversity on the level of cooperation for a variety of public goods games,  but not for  the snowdrift game \cite{Pena_2011,Pena_2016}.  In particular, $p_N$ in Eqs. (\ref{Fc}) and  (\ref{Fd})  is replaced by $ q_N = N p_N/m$, where $m$ is the mean group size. These different averages reflect different game scenarios. For example, in our scenario (see also \cite{Xu_2022}), a target player is randomly chosen from the infinite population and 
a value of group size $N$ is chosen from the distribution $p_N$.  The target player then participates in a game with other $N-1$ players randomly chosen from the infinite population. In the alternative scenario \cite{Pena_2011,Pena_2016}, the players are already segregated into groups of variable size, weighted by the distribution $p_N$, so that when we randomly choose a target player from the infinite population, we also choose the size of her playing group $N$. Thus, players in large groups are more likely to be selected as target players, hence the term $N p_N$ in $q_N$. The factor $m$ in the denominator of $q_N$ guarantees the normalization.  In subsection \ref{sec:R1} we show how our main results are  modified in this alternative scenario.

  An  advantage of  studying $N$-person games in a dynamic or static structured population scenario over
considering the  payoff accumulation of repeated $2$-person games (see, e.g., \cite{Meloni_2009}) is that there is no ambiguity in defining the payoff of an isolated player.

  \section{Results}\label{sec:res}

 In this section we consider the  annealed equilibrium solutions $x_{ann}^*$  and the equilibrium solution $x_m^*$ for a group of fixed size $m = \sum_{N=1}^\infty N p_N > 1$ for the three group-size distributions introduced in Section \ref{sec:casual}. Of course, for evaluating $x_m^*$ we must assume that the equations of Section \ref{sec:game} can be analytically continued to real $N$.  We find $x_{ann}^* \geq x_m^*$ regardless of the values of the parameters $\kappa$ and $r$ and of the group-size distribution, as expected \cite{Pena_2011}. Interestingly, for $x  \approx 0$  we can rewrite Eqs. (\ref{fc1}) and (\ref{fd1}) as
 \begin{equation}
 f_c(N,x) \approx b - c + \frac{c x}{2} (N-1)
 \end{equation}
 and
 \begin{equation}
 f_d(N,x) \approx b x (N-1),
 \end{equation}
 respectively. Carrying out the average over $N$ we find 
 \begin{equation}
 F_c(x) \approx b - c + \frac{c x}{2} (m-1)
 \end{equation}
 and
 \begin{equation}
 F_d(x) \approx b x (m-1).
 \end{equation}
Finally, equating these annealed payoffs yields the coexistence fixed point 
\begin{equation}\label{xan0}
 x_{ann}^* \approx 2 \frac{1-r}{m -1},
 \end{equation}
 which holds for any group-size distribution, provided that $ r \approx 1$. Comparing Eq. (\ref{xan0}) with the corresponding equation for a group of fixed  size, i.e., Eq. (\ref{xN0}), we conclude that $x_{ann}^* \approx x_m^*$ for $ r \approx 1$. 
 
 At this point, we can see that for games where the payoffs are linear functions of the number of cooperators, such as the $N$-person prisoner's dilemma,  we have $x_{ann}^* = x_m^*$ so that the group dynamics has  no significant  effect on the outcome of the game.
 

 \subsection{Zero-truncated Poisson distribution}
 
 Explicitly performing the sum over $N$ in Eq. (\ref{Fc}) using the zero-truncated Poisson distribution (\ref{pi_a0}) yields
 \begin{eqnarray}
 F_c(x)  & = &  b - \frac{c}{\kappa x} \sum_{N=1}^\infty \frac{a^N}{N!} \frac{1}{N} \left [ 1 - (1-x)^{N} \right ] \nonumber \\
  & = &  b - \frac{c}{\kappa x} \left [ g \left ( a \right ) - g \left ( a(1-x)  \right ) \right ] 
\end{eqnarray}
 where  $a= \ln (1 + \kappa )$ and
  \begin{equation}
 g(z) = \sum_{i=1}^\infty \frac{z^i}{i  i!}  =  \mbox{Ei} (z) - \gamma - \ln z.
  \end{equation}
 Here $\gamma$ is Euler's constant and $ \mbox{Ei} (z)$  is the exponential integral to be evaluated numerically  \cite {Harris_1957}. Instead, the  sum over $N$ in Eq. (\ref{Fd}) can be done in a closed form resulting in
 \begin{equation}
 F_d (x) =  b - \frac{b}{\kappa} \frac{1}{1-x} \left [ (1+\kappa)^{1-x} - 1 \right ] .
 \end{equation}
The coexistence fixed point is determined by the condition $ F_c(x_{ann}^*)  = F_d(x_{ann}^*) $. We can find an approximate analytical solution  for   $x_{ann}^* \approx 1$., viz.,
\begin{equation}
1-x^*_{ann}  \approx \left [ r - \frac{a}{g(a)}   \right ] \frac{1}{a/2 + a/g(a) - 1}  \frac{g(a)}{a},
 \end{equation}
which is valid if $r$ is close to  the critical parameter 
\begin{equation}\label{rcTP}
r_c = \frac{a}{g(a)} .
 \end{equation}
Note that  if $ r < r_c$, the all-cooperators  fixed point $x_{ann} = 1$ is  stable.  For $\kappa \to 0$ we have $a \approx \kappa $, so that  $r_c \approx 1 - \kappa/4$. For $\kappa \to \infty$ we have $a \approx \ln \kappa$, so that  $r_c \approx \ln^2 \kappa/\kappa$. This last result is easily obtained using the asymptotic expansion of the exponential integral \cite{Harris_1957}. So as $\kappa$ increases from $0$ to $\infty$, $r_c$ decreases from $1$ to $0$.

%
\begin{figure}[ht]
\centering
\includegraphics[width=1\columnwidth]{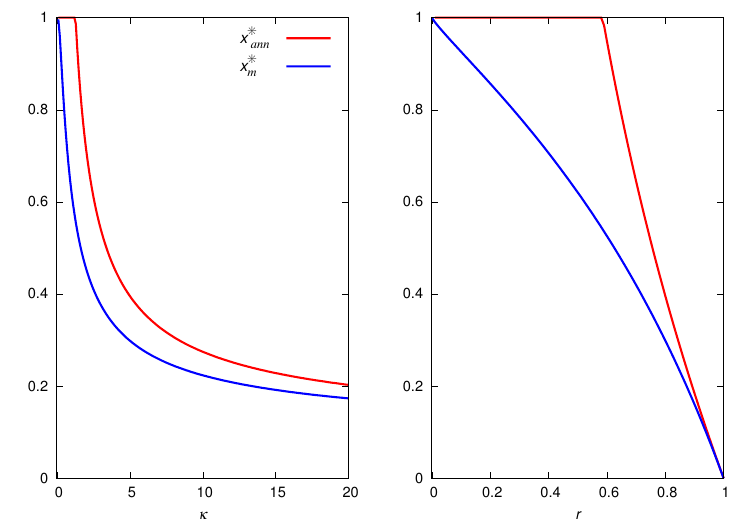}
\caption{Equilibrium cooperator  frequencies for the annealed  $x_{ann}^*$ and  mean group size $x_{m}^*$  scenarios as function of (left panel) the aggregation-disaggregation ratio $\kappa$ for $r=0.8$ and (right panel) the cost-benefit ratio $r$ for $\kappa=5$.  The group-size distribution is the zero-truncated Poisson distribution.
} 
\label{fig:1}
\end{figure}
%

Figure \ref{fig:1} shows the dependence of the equilibrium cooperator  frequencies on the parameters $\kappa$ and $r$. The striking effect of the group dynamics is to produce threshold phenomena such that for small $\kappa$ (i.e., small mean group sizes) or small $r$ (i.e., the cost of cooperation is small compared to the benefit of completing the task) the population is dominated by cooperators.
For $\kappa \to \infty$ we can again use  the asymptotic expansion of the exponential integral to show that $x^*_{ann} $ vanishes as
$y/\ln (\kappa)$ with $y=y(r)$ given by Eq. (\ref{y1}). Of course, $x_m^*$ shows the same asymptotic behavior, since Eq. (\ref{i_al0}) yields   
$m_P \approx \ln (\kappa)$ for the mean group size. So we have $x^*_{ann} \approx x_m^* \sim 1/\ln \kappa$ for large $\kappa$. In addition, 
$x^*_{ann} \approx x_m^*$ close to $r=1$ as shown by Eq. (\ref{xan0}).

%
\begin{figure}[t]
\centering
\includegraphics[width=1\columnwidth]{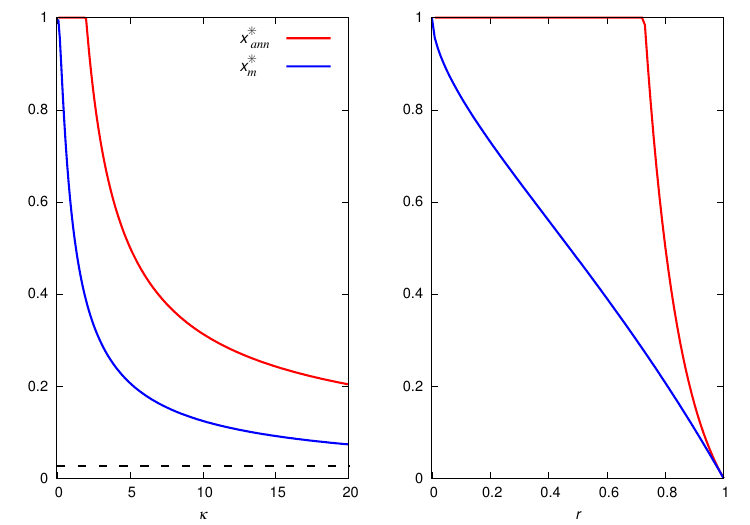}
\caption{Equilibrium cooperator  frequencies for the annealed  $x_{ann}^*$ and  mean group size $x_{m}^*$  scenarios as function of (left panel) the aggregation-disaggregation ratio $\kappa$ for $r=0.8$ and (right panel) the cost-benefit ratio $r$ for $\kappa=5$.  For $\kappa \to \infty$ we find $x_{ann}^* \approx 0.026$ (horizontal dashed line)  and  $x_{m}^* \to 0$  as $\ln \kappa/\kappa$. The group-size distribution is the logarithmic series distribution.
} 
\label{fig:2}
\end{figure}
%
 
 \subsection{Logarithmic series distribution}
 
 As before, we  carry out the sum over $N$ in Eq. (\ref{Fc}) using the logarithmic series distribution (\ref{pi_a1}) to obtain
 \begin{equation}
 F_c (x) = b -  \frac{c}{x\ln (1+\kappa)} \left [ \mbox{Li}_2 (u) - \mbox{Li}_2 \left ( u(1-x) \right ) \right ]
 \end{equation}
 where  $u= \kappa/(1 + \kappa )$ and
  \begin{equation}
 \mbox{Li}_2 (z) = \sum_{i=1}^\infty \frac{z^i}{i^2}  
  \end{equation}
  is the dilogarithm function \cite{Zagier_2007}, which must be evaluated numerically.  The  sum over $N$ in Eq. (\ref{Fd}) yields 
 \begin{equation}
 F_d (x)  = b + \frac{b}{\ln (1+\kappa)} \frac{1}{1-x} \ln [1 -u(1-x)] .
 \end{equation}
For $x_{ann}^* \approx 1$, the condition $ F_c(x_{ann}^*)  = F_d(x_{ann}^*) $ that determines 
the coexistence fixed point reduces to
\begin{equation}
1 - x_{ann}^* \approx  \left [ r - \frac{u}{\mbox{Li}_2(u)}  \right ]   
\frac{1}{u/2 +u/\mbox{Li}_2(u) - 1} \frac{\mbox{Li}_2(u)}{u},
 \end{equation}
which is valid when $r \approx r_c$ where
\begin{equation}\label{rcL}
r_c = \frac{u}{\mbox{Li}_2(u)} .
\end{equation}
For $r  < r_c$ the all-cooperators fixed point $x_{ann} = 1$ is stable. For $\kappa \to 0$ we have $u \approx \kappa$ so that $r_c \approx 1 - \kappa/4$ as  for the zero-truncated Poisson distribution. The  evaluation  of $r_c$ for $\kappa \to \infty$ is a bit more complicated.  In this case we have $u \approx 1 - 1/\kappa$ and so we need to find an approximation for $\mbox{Li}_2(1-z)$ for small $z$. This can be done using the identity \cite{Zagier_2007}
\begin{equation}
\mbox{Li}_2(z) + \mbox{Li}_2(1-z) = \frac{\pi^2}{6}  -\ln (z) \ln(1-z) ,
\end{equation}
which for small $z$ becomes 
\begin{equation}\label{liz}
 \mbox{Li}_2(1-z) \approx \frac{\pi^2}{6}  -z + z \ln (z) .
\end{equation}
Thus,  $r_c \approx 6/\pi^2 + (36/\pi^4) \ln \kappa/\kappa$. So as $\kappa$ increases from $0$ to $\infty$, $r_c$ decreases from $1$ to $6/\pi^2 \approx 0.608$. This means that $x_{ann}=1$ is stable for $r < 6/\pi^2$ regardless of the value of $\kappa$.

\begin{figure}
\centering
\includegraphics[width=1\columnwidth]{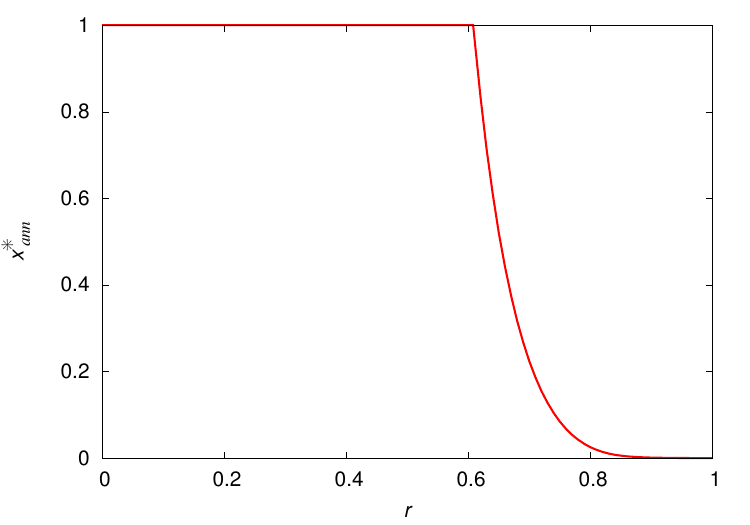}
\caption{Equilibrium cooperator frequency $x^*_{ann}$ in the limit  $\kappa \to \infty$ for group sizes distributed by the logarithmic series distribution. The all-cooperators fixed point  $x^*_{ann}=1$  is stable for $ r < 6/\pi^2$. 
} 
\label{fig:3}
\end{figure}

Figure \ref{fig:2} shows the dependence of the equilibrium cooperator frequencies $x_{ann}^*$ and $x_m^*$ on the parameters $\kappa$ and $r$. Since $x^*_m \sim 1/m $ for a group of large  fixed  size  $m$ we have $x_m \sim \ln \kappa/\kappa$ for large $\kappa$. Most interestingly, however, the annealed fixed point  $x_{ann}^*$ tends to a nonzero value in  the limit $\kappa \to \infty$. In fact, as shown before
 $x_{ann}^* = 1$ for $r < 6/\pi^2$ even in this limit. For $r > 6/\pi^2$,   the equation  for the coexistence fixed point, viz. $ F_c(x_{ann}^*)  = F_d(x_{ann}^*) $,    becomes
\begin{equation}
 - x_{ann}^* \ln x_{ann}^* =r (1-x_{ann}^*) \left [ \frac{\pi^2}{6} - \mbox{Li}_2 (1-x_{ann}^*) \right ] ,
\end{equation}
where we have set $u=1$  (i.e., $\kappa \to \infty$). Figure \ref{fig:3} shows the solution of this equation	as a function of the cost-benefit ratio. Note that although the mean group size diverges for $\kappa \to \infty$, the frequency of cooperators in the population is non-zero in stark contrast to the results for a fixed (large) group size. Use of Eq. (\ref{liz}) yields $x_{ann}^* \approx \exp [ -1/(1-r) ]$ for $r \approx 1$.

%

 \subsection{Isolate-pair Bernoulli distribution}
 
 In this case, only isolates ($N=1$) and pairs ($N=2$) are present in the population in the proportions $p_1$ and $p_2$ given by Eqs. (\ref{p1i}) and (\ref{p2i}). For the annealed average we have
 \begin{equation}
 F_c(x) = b - c + \frac{p_2 c}{2} x
\end{equation}
and 
 \begin{equation}
 F_d(x) = p_2 b x.
\end{equation}
The condition $ F_c(x_{ann}^*)  = F_d(x_{ann}^*) $ yields 
 \begin{equation}\label{xB}
x_{ann}^*  = \frac{1}{p_2} \frac{1-r}{1-r/2}
\end{equation}
for $p_2 >  (1-r)/(1-r/2)$. Otherwise, the fixed point $x_{ann}=1$ is stable. Thus, the cooperators dominate the population for $r \leq r_c$, where
\begin{equation}\label{rcB}
r_c = \frac{1}{1 + \kappa/4} .
\end{equation}
 As before, $x_m^*$ with $m=m_B$ given by Eq. (\ref{i_ali}) must be found by solving numerically Eq. (\ref{fc=fd}). 
For $\kappa=2$ we have $p_1=p_2 = 1/2$ and  so Eqs. (\ref{xB}) and (\ref{rcB}) give the solution for  the uniform group size distribution  with $N_m=2$ \cite{Xu_2022}.

\begin{figure}[h]
\centering
\includegraphics[width=1\columnwidth]{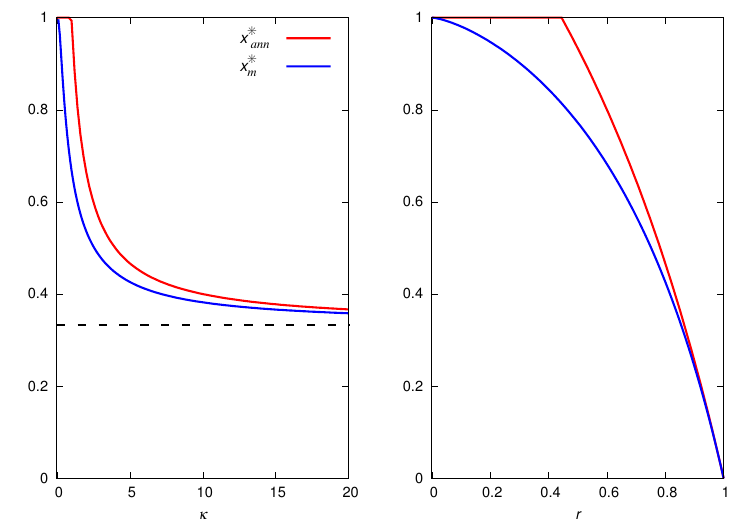}
\caption{Equilibrium cooperator  frequencies for the annealed  $x_{ann}^*$ and  mean group size $x_{m}^*$  scenarios as function of (left panel) the aggregation-disaggregation ratio $\kappa$ for $r=0.8$ and (right panel) the cost-benefit ratio $r$ for $\kappa=5$. The horizontal dashed line indicates the equilibrium frequency for $\kappa \to \infty$, viz.,  $x_{ann}^*= x_{m}^*= 1/3$. The group-size distribution is the isolate-pair Bernoulli distribution.
} 
\label{fig:4}
\end{figure}

Figure \ref{fig:4} summarizes the main results for the isolate-pair Bernoulli distribution of group sizes. For $\kappa \to \infty$ we have $p_1=0$ and $p_2=1$. Thus, this limit corresponds to a group of fixed size $N=2$ and so  we have $x_{ann}^*= x_{m}^* = x_2^* = (1-r)/(1-r/2) $.

%
%

\subsection{Threshold parameters}\label{sec:R1}

As observed before, the  unexpected effect of the group dynamics is the stabilization of the all-cooperators fixed point for $r  < r_c ( \kappa)$.
This happens only because the equilibrium group-size distributions resulting from the casual group dynamics allow for isolates, as in the case of the uniform distribution \cite{Xu_2022}. In fact, if only groups of size $N>1$ are allowed then the all-cooperators regime can never arise, since a single defector will have payoff $b$ that is higher than the payoff of a cooperator, regardless of the size and composition of the play group \cite{Xu_2022}.

\begin{figure}
\centering
\includegraphics[width=1\columnwidth]{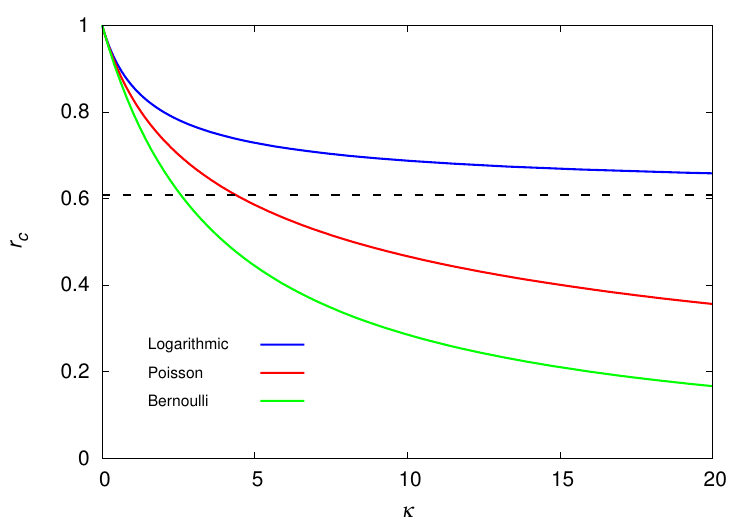}
\caption{Critical value of the  cost-benefit ratio $r_c $ below which the all-cooperators fixed point is stable as function of the aggregation-disaggregation ratio $\kappa$ for the  isolate-pair Bernoulli, zero-truncated Poisson and logarithmic series distributions, as indicated. The horizontal  dashed line is the value $r=6/\pi^2$ below which the all-cooperators fixed point is stable for the   logarithmic series distribution.
} 
\label{fig:5}
\end{figure}

   Figure \ref{fig:5} shows the  critical value of the cost-benefit ratio against the  aggre\-ga\-tion-dis\-aggre\-ga\-tion ratio $\kappa$. We recall that $r_c \approx 1 - \kappa/4$  for  small $\kappa$ for the three group-size distributions. For large $\kappa$,  we found $r_c \sim 1/\kappa$ for the isolate-pair Bernoulli distribution and  $r_c \sim \ln \kappa /\kappa$ for the zero-truncated Poisson distribution. Surprisingly, for the logarithmic series distribution $r_c$ does not vanish in the limit $\kappa \to \infty$. Instead, $r_c$ tends to the finite value $1/\mbox{Li}_2(1) = 6/\pi^2$, which signals that $x_{ann}=1$ is stable for $r <  6/\pi^2$ regardless of the value of $\kappa$. Note that for the uniform group-size distribution, $r_c$ vanishes when the upper bound  $N_m$  diverges \cite{Xu_2022}.    Recent empirical data on  human face-to-face interactions   suggest that group-size distributions are fat-tailed \cite{Cattuto_2010,Starnini_2016}. In this sense,  the logarithmic  series distribution is a more realistic model than the zero-truncated Poisson and the uniform distributions.  We expect that a fat-tail group-size distribution would  further promote the all-cooperators regime.

 Equations   (\ref{rcTP})   and   (\ref{rcL}) suggest  that there is a simple general expression for the critical  parameter $r_c$ in terms of the group size distribution $p_N$. In fact, since the all-cooperators fixed point is locally stable for $F_c(1) - F_d(1)> 0$,  the critical  parameter $r_c$ is determined by the condition $F_c(1) = F_d(1)$, which is simply rewritten as 
 \begin{equation}\label{harm}
 \frac{1}{r_c} = \frac{1}{p_1} \sum_{N=1}^\infty p_N \frac{1}{N},
 \end{equation}
 where we have taken the limit $x \to 1$ in Eqs. (\ref{fc1}) and (\ref{fd1}). It is clear from this equation that $p_1 > 0$ is a sufficient condition for $r_c > 0$ and thus for the existence of an all-cooperators regime. Equation (\ref{harm}) has been derived in a much less transparent way for the uniform distribution and then assumed to hold true for non-uniform distributions as well \cite{Xu_2022}. Our elementary derivation of Eq. (\ref{harm}),  which is essentially the condition for  the stability of the all-cooperators fixed point, proves that it is indeed valid in general. 
 
If we consider an alternative game scenario where players in large groups are more likely to be selected as target players  \cite{Pena_2011,Pena_2016}, i.e., $p_N$ is replaced by $q_N=Np_N/m$ in   Eqs. (\ref{Fc}) and  (\ref{Fd}), then we simply  get    $r_c = p_1$. This shows that the  all-cooperators regime is also stable in this scenario, provided that $r < r_c$. However, since   $p_1$ vanishes as $1/\ln \kappa$ for the logarithmic series distribution, we have $r_c \to 0$ in this case.

%
%

 \section{Discussion}\label{sec:conc}

It is instructive to note the strong, but  largely overlooked, similarity between  the replicator equation approach to public goods games and the trait group framework proposed by Wilson in the late 1970s \cite{Wilson_1975,Wilson_1980,Okasha_2009} to  study traits that affect not only the fitness of the individuals that exhibit them, but also that of other members of the population. Examples of such traits are behaviors that alter the environment (e.g., pollution and resource depletion) and thus ultimately affect the entire population, which is precisely the subject of public goods games. In addition, a key component of Wilson's trait group idea is that while the fitness of individuals is determined locally, taking into account interactions within their trait groups, their chances of reproduction are dictated by competition in the population at large. This is the same procedure that justifies the use of the replicator equation in $N$-person games: a player's payoff is determined by the composition of the group of $N$ players to which she belongs, but the decision to change her strategy or not is determined by comparison with the payoff of another player randomly selected from the population, who most likely does not belong to the focal player's group \cite{Hannelore_2010,Zheng_2007}. For the $N$-person snowdrift game, the payoff (or fitness) for a cooperator in a group with $i>0$ cooperators, Eq. (\ref{Pci0}),   is rewritten  in the trait group  formulation  as 
\begin{equation}
P_c(i) = b - c + (i-1) \frac{c}{i},
\end{equation}
where $b-c$ is the effect of a cooperator on herself and $c/i$ is the effect of the other cooperators \cite{Wilson_1980}. When $b < c$ (i.e., $r >1$), so-called strong altruism, the cooperator confers a fitness benefit on other members of her group, but suffers a selective disadvantage herself relative to the baseline fitness of an isolated defector. Regardless of the group-size distribution, cooperators in this scenario are doomed to extinction \cite{Hamilton_1975,Charlesworth_1979},  unless there are some strong positive assortment between them \cite{Hamilton_1964}. However, when $b>c$ (i.e., $r < 1$), so-called weak altruism, the maintenance of cooperators in the population is possible given sufficient variability among trait groups. As for the $N$-person games, see Eqs.  (\ref{fc1}) and  (\ref{fd1}), the binomial distribution is the usual choice to  model the composition  of the  trait groups \cite{Alves_2000}.  Although game theory, and in particular the $N$-person prisoner's dilemma, has already proven useful in integrating elements (e.g., Hamilton's rule and the Price equation) of multilevel selection and inclusive fitness theories for the evolution of altruism \cite{Fletcher_2007}, the connection between the replicator equation and Wilson's trait group framework seems to go unnoticed.  It might be claimed that neither our model of game dynamics in casual groups, nor Wilson's trait group model, are models of group selection sensu stricto \cite{Maynard_1976}, which require birth and death processes at both the group and individual levels (see, e.g., \cite{Simon_2013,Fontanari_2014,Luo_2017,Cooney_2020}). However, assuming that differential extinction and colonization of groups are prerequisites of real group selection is not without caveats.  As discussed by Okasha  \cite{Okasha_2009} at some length, making extinction and colonization processes as isomorphic to the birth and death processes of individual organisms can be an illusion because ultimately, in most group models, the focal units are the individuals, not the groups (see also \cite{ Heisler_1987}, p. 584).

A fundamental result of Wilson's approach is that cooperation is promoted by increasing the variability of group composition or, if fitness is not a linear function of the number of cooperators, by increasing the variance of a cooperator's fitness.  Direct application of the variance decomposition formula \cite{Weiss_2005}  shows that a non-degenerate distribution of group sizes  always increases the  variance of a  cooperator's fitness. Of course, this qualitative understanding is no substitute for the detailed analysis obtained by solving the replicator equation and, in particular, cannot predict the existence of an all-cooperators  regime.

For a fixed finite group size, the original $N$-person snowdrift game already guarantees the maintenance of a certain fraction $x_N^*$ of collaborators in the infinite population at equilibrium. This proportion vanishes as $x_N^* \sim 1/N$ as $N$ increases \cite{Zheng_2007} (see also \cite{Pena_2018}). This observation adds value to our finding that for the logarithmic series distribution, the fraction of cooperators $x_{ann}^*$ does not vanish as the mean group size becomes arbitrarily large (i.e., in the limit where the aggregation-disaggregation ratio $\kappa$ goes to infinity), as shown in Fig.\ \ref{fig:3}.  In contrast, $x_{ann}^* \sim 1/\ln k \to 0$ for the zero-truncated Poisson distribution in this limit. A possible reason may be that the fraction of isolates vanishes much slower for the logarithmic series distribution ($p_1 \sim 1/\ln k$) than for the zero-truncated Poisson distribution ($p_1 \sim \ln k/k$), and being isolated is advantageous for cooperators since $b>c$. 
The  $\kappa \to \infty$ limit corresponds to the situation where once an isolate joins a group, it stays in that group forever, so in that sense the group dynamics is effectively  frozen after the isolates disappear.  In this case, our results indicate that there is a transition between the regime where defectors dominate the population (i.e., $x_{ann}^* = 0$) and the coexistence regime (i.e., $x_{ann}^* > 0$) as group attractiveness increases from $\alpha=0$ to $\alpha=1$ (see Section \ref{sec:casual}). 
Overall, our conclusions dovetail with Wilson's general argument above that a structured population is beneficial for individuals carrying a weak altruistic trait \cite{Wilson_1975} as well as with several studies on dynamic grouping (see below).

To conclude, we must mention some noteworthy previous studies on dynamic grouping in the $N$-person snowdrift game. For instance,  this problem has been  addressed using a group dynamics originally proposed to model the herd behavior responsible for the fat-tail distribution in returns of financial price data \cite{Ji_2011}. However, the use of the same time scale for the imitative dynamics that agents use to update their strategies and for the group dynamics, as well as the complexity of the aggregation-disaggregation process, precludes an analytical study based on the replicator equation. Nevertheless, the simulations show that with an appropriate choice of parameters, the population can reach a fully cooperative state \cite{Ji_2011}. 
In the same vein, the use of the uniform distribution of group sizes, which has no biological or sociological justification, yields results very similar to ours, as we have already pointed out \cite{Xu_2022}: the presence of isolates guarantees the stability of an all-cooperators regime. Again, because imitation and group dynamics have the same time scale, the framework of the replicator equation does not apply directly, and more complicated analytical tools are needed to derive simple results such as Eq. (\ref{harm})  \cite{Xu_2022}.  It is also imperative to mention studies on the fluctuating group size  of a  continuous version of the snowdrift  game \cite{Doebeli_2004}, where  the level of cooperative investment is not restricted to two values only (i.e., $0$ or $b$), but can take on a continuum of values and, more importantly,  can vary across  players   \cite{Brannstrom_2011}. 

Two other interesting approaches to promoting cooperation in $N$-person public goods games are success-driven migration, in which individuals consider several alternative locations within a given migration range and choose the one that is most favorable to their strategy \cite{Helbing_2009}, and conditional dissociation, in which individuals have the option of leaving their partners in response to their behavior \cite{Izquierdo_2014}.
Similarly, the group sensitivity approach assumes that in addition to the cooperation or defection strategies, each player has a preferred group size that is also changed by the imitation dynamics \cite{Shi_2013} (see \cite{Powers_2011} for a genetic version of this approach).  Of course, success-driven migration,  conditional dissociation and group sensitivity  are particular realizations of  positive assortment between cooperators. Since in our model the aggregation and disaggregation parameters are the same for all individuals, we do not consider any kind of assortment mechanism.  In addition, these approaches are based on agent-based simulations and lead, as expected, to an increase in the frequency of cooperators.

Our approach is unique in the sense that it is based on a classical sociological model of group formation \cite{Coleman_1961,White_1962} and allows a full analytical solution of the equilibrium state of the population of players within the framework of the replicator equation. These ingredients allow us to show that cooperation persists even when the average group size becomes arbitrarily large, provided that players, regardless of their strategies, are more attracted to large groups.

\bigskip

\acknowledgments
JFF is partially supported by  Conselho Nacional de Desenvolvimento Ci\-en\-t\'{\i}\-fi\-co e Tecnol\'ogico  grant number 305620/2021-5. MS is funded by  grant PID2021-127107NB-I00 from Ministerio de Ciencia e Innovaci\'on (Spain),  grant 2021 SGR 00526 from Generalitat de Catalunya, and the Distinguished Guest Scientists Fellowship Programme of the Hungarian Academy of Sciences  (https://mta.hu).

\end{document}